\documentstyle[12pt]{article}
\def\be{\begin{equation}}
\def\ee{\end{equation}}
\def\bea{\begin{array}}
\def\eea{\end{array}}

\begin{document}
\pagestyle{plain}
\vskip .25in

\begin{flushright}
{hep-th/9610181}\\
{IPM-96-168}
\end{flushright}

\begin{center}
\Large POINT-LIKE STRUCTURE IN STRINGS AND\\ 
\begin{center}
NON-COMMUTATIVE GEOMETRY
\end{center}
\small
\vskip .15in
by\\
\vskip .15in

Farhad Ardalan\footnote{E-mail: ardalan@theory.ipm.ac.ir}$\;\;$ and$\;\;$ 
Amir H. Fatollahi\footnote{E-mail: fath@theory.ipm.ac.ir}\\
\-

{\it Institute for Studies in Theoretical Physics and Mathematics (IPM),}\\ 
{\it P.O.Box 19395-5531, Tehran, Iran}\\
{\it and}\\
{\it Department of Physics, Sharif University of Technology,}\\
{\it P.O.Box 11365-9161, Tehran, Iran}\\

\vskip 2 in
{\bf ABSTRACT}
\end{center}

Dynamics of a free point particle on a multi world-line is presented
and shown to reduce to that of a bosonic string theory at the appropriate
limit. Other higher dimensional extended objects are argued to appear 
at other regions of the space of configurations of the theory.

\vskip .10in

\newpage
\setcounter{page}{2}
\pagestyle{plain}
\section{Introduction}
In the dramatic developments in the understanding of the strong coupling
limit of string theory of the last two years, two ideas have resurfaced:
Point-like structures in the small distance regime; and noncommutative 
aspects in these scales.

Point-like substructure of string theory has long been suspected and recently
extensively studied \cite{1,2}. More directly relevant to the recent 
developments of string duality are the realization that D0-branes play an
important role in string dynamics and in the presumably more fundamental
M-theory \cite{3,4,5}. In the distances shorter than string scale \cite{6}
there are signatures of the breakdown of the smooth structure of the world-sheet
\cite{4,7}; higher dimensional D-branes have been argued to be bound states
of D0-branes \cite{8}; and it has been conjectured that M-theory is matrix
model theory of D0-branes \cite{9}. 

The noncommutative structures of space-time in the Planckian scales have been
considered in various contexts; however, in relation to duality and D-branes
of string theory, an explicit realization of a noncommutative space-time
was given for the bound state of parallel D-branes \cite{10}; and is an 
important ingredient in the matrix model proposed for M-theory \cite{9}.

Although this approach to noncommutative  structure  of space-time is 
distinct \cite{5}
from the celebrated mathematical formulation of noncommutative geometry by
A. Connes \cite{11} as e.g. applied to the standard model of electro-weak 
interactions (though may be dual to it), it is however tempting to apply 
Connes' methods to problems of string theory. Also in the context of string  
field theory noncommutative geometry has been utilized \cite{12}. Aside 
from the standard model, these methods have been used 
in a number of areas \cite{13,14}.

As a first step in this project we try to show that the generalized  
geometrical concepts (such as manifold, differential structure, metric, 
distances,...) of noncommutative geometry can be used as a 
natural framework for studying substructure of strings. 
We consider a point particle on a 
specific multi- world-line manifold and formulate its dynamics in Connes' 
framework. In a certain limit of a particular region of the space of
configurations of the theory we recover the usual bosonic string theory.
We argue that  higher ($p$) dimensional extended objects are also derivable 
from other regions of the configuration space. 

We must emphasize that although the space-time coordinates of our 
particle are not strictly speaking noncommutative (they are diagonal matrices),
yet by putting a noncommutative differential structure on the geometric space,
we may find non-trivial results; it is the application of the 
general noncommutative framework of Connes 
to the simplest example of a free particle that is our main interest here which 
we hope to generalize to more complicated cases.  

The advantages of our approach to discretization of strings are 
as follow: 

1- It is manifestly relativistically covariant (no need to go to the light cone
gauge).
 
2-There is no need for an ad hoc potential for the inter-particle interactions.
The required  potential is a consequence of the natural differential 
structure put on the geometric space. 

3-The Virasoro constraints, which signal the enhancement of  
conformal symmetry in the continuum limit of the discrete strings, follow 
directly from the formalism.

\section{Formalism}
To begin with let us recall the action of a free relativistic particle,
\be
S=-m \int d\tau \sqrt{-{\dot{x}}^2} ,
\ee
\noindent where $\tau$ is a lable for 
the world-line. From the definition of momentum, 
$p_\mu=\partial L / \partial \dot{x}^\mu$, the constraint,
\be 
p^2+m^2=0                               ,
\ee
follows. The Hamiltonian of the system is identically zero,
\be
H_0=p.\dot{x} -L=0                              .
\ee

 For our purposes, it is convenient to linearize the action (1) by 
introducing a metric on the world-line and obtain,
\be 
S=1/2 \int\;e\; d\tau (e^{-2}\; {\dot{x}}^2 -  m^2),
\ee
\noindent where $e^{-2}$ is the single component of the metric
($g^{\tau \tau}$). The equation of motion for $e$ yields, 
\be
-\partial L/\partial e= e^{-2}{\dot{x}}^2 +m^2=0 ,
\ee
\noindent which is interpreted as a constraint. 

Let us reformulate the above in terms of Connes' noncommutative geometric
(ncg) framework. In this framework a manifold, in this case the world-line of 
the particle, ${\cal (M)}$ is replaced by an algebra ${\cal A}$, 
(in general the algebra of the smooth functions on the manifold, as it can 
be shown that the space of smooth functions on a manifold, uniquely determines 
the manifold). The differential structure is given by a Dirac operator $D$ 
acting on a Hilbert space 
${\cal H}$. In the simple case we choose it to be
${\cal L}^2({\cal M})$, the set of all square integrable 
functions on
${\cal M}$, on which a representation of the algebra also acts.
All the calculations are to be understood in the sense of the chosen Hilbert 
space.
In the above 
case of an ordinary particle, the algebra is simply the algebra of functions 
on the real line,

\be 
{\cal A}=C^\infty {\cal (M)}                                    ,
\ee
\noindent and the natural Dirac operator is

\be 
D=i e^{-1} \partial_\tau                           ,
\ee

\noindent  here $e$ is the one dimensional position dependent Dirac 
$\gamma$-matrix ( obeying the Clifford algebra 
$[e^{-1},e^{-1}]_{+}=2 e^{-2}= 2 g^{\tau \tau}$).
 The coordinates of a particle are functions (scalars) in 
the sense of world-line manifold; thus they belong 
to the algebra ${\cal A}$. So we introduce the  $x^\mu(\tau)$'s as coordinates, also as representation of ${\cal A}$ which are operators
on the Hilbert space ${\cal H}$. Then, the action (4) in the language of noncommutative 
geometry becomes,

\be
S=\kappa\; Tr_\omega \biggl( ([D,x^\mu][D,x_\mu]^*-m^2) \mid D\mid ^{-1} \biggr),
\ee
\noindent where $\kappa$ is a constant and $Tr_\omega$ is the Dixmier trace 
defined as below for a (compact) operator $T$
$$
 Tr_\omega (T)= lim_{N \rightarrow \infty}  
 {1\over{log N}} \sum_{j=1}^{N}(t_1+t_2+...+t_j)
 $$
and $t$'s are absolute values of the eigenvalues of $T$ arranged in a 
decreasing sequence.
One can show that (8) is equivalent to (4) \cite{11,13}.
                                
We will now present our generalisation of the free particle which as we will 
show is capable of describing discretized extended objects, in particular
descretized strings. We take the more general algebra
 \be
{\cal A}=C^\infty{\cal (M)}\oplus C^\infty{\cal (M)} \oplus ... \oplus C^\infty{\cal (M)}=C^\infty{\cal (M)} \otimes Z_N
\ee

\noindent which is geometrically equivalent\footnote{More precisely they are  
topologically equivalent, so one can, as we will do, put different metrics
on them.}
with $N$  copies of 
the manifold $\cal M$ (in our case world-line manifold). For its 
representation we take,

\be
X^ \mu= diag\;  (x_1^ \mu( \tau),x_2^ \mu( \tau),...,x_N^ \mu( \tau))
\ee

What is nontrivial is the choice of the Dirac operator which is an
$N \times N$ hermitian matrix. At a particular region of the possible 
'configurations' of this matrix we take 
\footnote{A more general form of Dirac operator in 1-dimension is 
$D=e^{-1} \; i\;\partial_\tau\;+\;\Phi(\tau)$ where $e^{-1}$ is the inverse
of the matrix (13) and $\Phi(\tau)$ is a hermitian $N\times N$ matrix.},

\be
D=\left( \matrix{ 
ie_1^{-1}\partial_\tau & \Phi/\sqrt{2} & 0 & & \cdots & & 0 \cr 
\Phi/\sqrt{2} & ie_2^{-1}\partial_\tau & \Phi/\sqrt{2} & 0& \cdots & & 0 \cr
0 & \Phi/\sqrt{2} & ie_3^{-1}\partial_\tau & \Phi/\sqrt{2} & 0 &\cdots & 0 \cr
& & &  \ddots  & & \cr  
0 & \cdots & & & 0 &  \Phi/\sqrt{2} & ie_N^{-1}\partial_\tau }\right)
\ee

\noindent or,

\be
D_{ij}= i e_i^{-1} \partial_\tau \delta_{ij} + {\Phi \over \sqrt2}\; 
( \delta_{i,j-1} + \delta_{i,j+1})                                                 ,
\ee

\noindent where $\Phi$ is a constant. Defining a generalized determinant of
the metric \cite{14} as,
\be
e=diag\;  (e_1,e_2,...,e_N)                                          ,
\ee

\noindent one can write the action in the form similar to that of the standard
free particle eq.(8)

\be
S= 1/2 \; \int d\tau\; tr \biggl( e\;([D,X^\mu][D,X_\mu]^*-m^2 {\bf 1}_N)\biggr)     ,
\ee
\noindent or explicitly as,

\be
S=1/2 \int d\tau \sum_{j=1}^{N} \biggl( e_j ^{-1} {\dot{x}}_j{^2}- e_j 
{\Phi^2 \over 2} \;(x_{j,j-1}^2 + x_{j,j+1}^2 + m^2)\biggl)
\ee
\noindent where $x_{k,l}=x_k-x_l$.

Let us now restrict ourselves to the case of open chains and we use following 
boundary conditions (the same which one put for open strings),

\be
x_0-x_1=x_N-x_{N+1}=0           ;
\ee
\noindent closed chains can be similarly treated.

The equation of motion for $e$'s is interpreted as constraints,

\be
T_j \equiv - \partial L/ \partial e_j =e_j^{-2} {\dot{x}}_j{^2} + 
{\Phi^2 \over 2}\; x_{j,j-1}^2 + {\Phi^2 \over 2}\; x_{j,j+1}^2 + m^2=0 .
\ee

By solving for $e$'s and inserting in the action we find the Nambu-Goto like 
action,

\be 
S=\;-\;\int d \tau \sum_{j=1}^N \sqrt{-{\dot{x}}_j{^2} 
 \biggl({\Phi^2 \over 2}\; (x_{j,j-1}^2 + x_{j,j+1}^2 )+m^2}\biggr)            .
\ee

The momenta,
  
\be
p_{j \mu} \equiv \partial L/ \partial {\dot{x}}_j{^\mu} =
{{\dot{x}}_{j\mu} \sqrt{{\Phi^2 \over 2}\; (x_{j,j-1}^2 + x_{j,j+1}^2 )+m^2} 
\over \sqrt{-{\dot{x}}_j{^2}}}                                     ,
\ee
\noindent now satisfy the constraints,

\be
p_j^{2}+{\Phi^2 \over 2}\; (x_{j,j-1}^2 + x_{j,j+1}^2 )+m^2=0       ,
\ee
\noindent which are the same as in eq.(17). The Hamiltonian of the
system is again identically zero; we therefore follow Dirac's recipe for a
constrained system.  First we introduce a set of functions $\lambda_n$ as a
linear combination of constraints $T_j$ of eq.(17). For convenience we set 
$m=0$,

\be
\lambda _n=(1/N)\sum_{j=1}^{N} cos\biggl({n\pi(j-1/2)\over N}\biggr)\; T_j\; ,
\ee

$$
0 \leq n \leq N-1.
$$

Because of the vanishing of the Hamiltonian, the Poisson brackets are satisfied
and we only have to ensure closure of the algebra of constraints. 
Now the total Hamiltonian is
\be
H=H_0+\;v_n \lambda_n+\;u_m \eta_m                                   ,
\ee
where $\eta$'s are secondary constraints and $v$'s and $u$'s are arbitrary 
constants which we choose them as follows:
\be
u_m=0  ,\; v_n={1\over2} \;\delta_{n0};
\ee
\noindent so we have
\be
H=1/2 \; \lambda_0= 1/2 \sum_{j=1}^{N}(p_j^2+ \Phi^2 \;x_{j,j-1}^2 ),
\ee

The equations of motion,
\be
{\dot{x}}_j^\mu=[H,x_j^\mu]_{PB},
\ee
\be
{\dot{p}}_j^\mu=[H,p_j^\mu]_{PB} ;
\ee
\noindent lead to,
\be
{\ddot{x}_j^\mu}+\; \Phi^2\; (2 x_j^\mu-x_{j+1}^\mu-x_{j-1}^\mu)=0,
\ee
\noindent with the open boundary conditions, $x_0=x_1,\; x_{N+1}=x_N$.

The solution is ,
\be
x_j^\mu(\tau)=x_0^\mu +{V_0^\mu \tau \over N+1} +{i\over \sqrt{N+1}} \sum_{n\neq 0}
{1\over {\omega_n}} \; \alpha_n^\mu e^{-i\omega_n \tau} cos\biggl({n\pi (j-1/2)\over N}\biggr),
\ee

$$
1\leq j \leq N
$$

\noindent  and satisfying $[x_j^{\mu},p_k^\nu]_{PB}=-\delta_{jk} \eta^{\mu\nu}$
yeilds the Poisson brackets,
\begin{eqnarray}
[\alpha_n^\mu,\alpha_m^\nu]_{PB}=i\delta_{m,-n} \eta^{\mu\nu}\omega_n,\nonumber \\
\mbox{} [x_0^{\mu},V_0^\nu]_{PB}=-\eta^{\mu\nu} ,  \nonumber  \\
\mbox{} [\alpha_n^\mu, V_0^\nu]_{PB}=[\alpha_n^{\mu},x_0^{\nu}]_{PB}=0,\nonumber \\
\omega_n=2 \Phi sin({n\pi\over {2N}}).
\end{eqnarray}

Then,
\begin{eqnarray}
\lambda_n=\biggl( 1+cos({n\pi\over {2N}})\biggr)\sum_m(\alpha_m.\alpha_{-m-n}+
\alpha_m.\alpha_{-m+n}) \nonumber  \\
\;+\;\biggl(1-cos({n\pi\over {2N}})\biggr)\sum_m(\alpha_m.\alpha_{m-n}+\alpha_m.\alpha_{m+n})
\end{eqnarray}

$$
0 \leq n \leq N-1
$$

The closure of the constraints for finite N is a complicated problem. 
We will instead study a simple limit. As is well known \cite{11,13} , in ncg the 
inverse of the off diagonal elements of the Dirac operator, ( note that their 
dimensions are $lenght^{-1}$), estimate the distance between the
individual parts of the geometric space, which in our case becomes the
 distance between the nearby world-line manifolds. We can therefore define the 
following as the continuum limit ( joining many world-line manifolds to produce 
a higher dimensional manifold),
$$
N\rightarrow\infty\;,\;\;\Phi \rightarrow\infty \;\;\; ,\Phi/N=1/\pi
$$

Then eq.(30) becomes (for large $N$ and long wavelenghts with respect
to $\Phi^{-1}$, $n \ll N$ ) 
\be
\lambda_n=L_n  \;+ \;L_{-n},
\ee

$$
0 \leq n
$$

\noindent where L's are Virasoro constraints ($\alpha_0 \sim V_0$). When we
take the Poisson bracket of  
$\lambda_n,\; (n\neq0)$ with $\lambda_0$, 
we find the secondary constraints,
\be
\eta_n=[\lambda_n,\lambda_0]_{PB}= \;i n(L_n-L_{-n}).
\ee

Combining (31),(32) we find all the Virasoro constraints as the constraints of
our system,
\be
L_n=0\;, \forall\; n\in Z \kern -.5em Z.
\ee
\noindent which obey the classical algebra
$$
[L_{m},L_{n}]_{PB}=i(m-n) L_{m+n}
$$

The continuum limit of (27) , where the term in the parenthesis divided by $\Phi^{-2}$
acts as the  second derivatives, is
\be
{\ddot{x}}_ \mu -{x_\mu}'' =0,
\ee
\noindent and the corresponding action is:
\be
S=1/2\;\int\;({\dot{x}}^2-{x'}^2)\;d\tau d\sigma,
\ee
\noindent and because $N.\Phi^{-1}=\pi, \sigma \in [0,\pi]$.

The action (35) with the Virasoro constraints, which appeared  as the natural
constraints of the dynamical system, is equivalent to the
classical Nambu string.

  In the above description we have not specified the nature of the target 
space as the intrinsic differential structure we have built on the world-line
does not require any further specification than that the 'coordinates' $x_\mu$
must represent the algebra. We think this is an advantage of the
noncommutative formulation. Yet in the spirit of the original application of
these methods to the standard model \cite{13}, we may interpret the formalism 
as a multi-layered world-line or multi-layered target space. The supersymmetric
string is also easily derivable from a similar noncommutative supersymmetric
free particle and will be presented elsewhere.

   This formalism also allows description of higher ($p$) dimensional 
   extended objects. In fact the Dirac operator of the theory has a large
number of possible configurations over which one has to sum in a quantum
path integral, of which that of eq.(12) which led to the string is only one
example. 

To get other higher ($p$) dimensional extended objects we must allow for other 
non-zero entries in the 
matrix for the Dirac operator. Depending on which entries are set equal to zero
different objects, including $p$ dimensional extended objects appear at large 
$N$ limit. For example
an appropriate\footnote{In general allowing the element $D_{jk}$ of   
the Dirac operator to be non-zero gives a link between the sites $j$ and $k$ 
of the lattice.} 
set of non-zero off diagonal constant matrix elements of $D$ 
results in the action for a $p$ dimensional extended object,

\be 
S=1/2 \int d\tau \sum_{j_1,...,j_p=1}^{N_1,...,N_p} 
\biggl(e_{j_1...j_p} ^{-1} {\dot{x}_{j_1...j_p}}^2- e_{j_1...j_p} 
{\Phi^2 \over 2} \;
\sum_{i=1}^p(x_{j_i,j_i-1}^2 + x_{j_i,j_i+1}^2 )\biggr) ,
\ee
\noindent where $x_{j_i,j_i\pm 1}=x_{j_1...j_i...j_p}-
x_{j_1...j_i\pm 1...j_p}$ and $N=N_1N_2...N_p$. Here $j_1...j_p$ are an 
appropriate one-to-one map from $Z \kern -.5em Z$ to $Z \kern -.5em Z^p$.
Again equations of motion of $e_{j_1...j_p}$ 's
give $N$ constraints $T_{j_1...j_p}$ .

Its continuum limit is,

\be 
S=1/2\;\int\;(e^{-1}\;{\dot{x}}^2-e\;\sum_{i=1}^p{x,_i}^2)\;d\tau d^p\sigma,\;\;
\ee

$$
\sigma _i\in [0,\pi],
$$

\noindent where $x,_i=\partial x/\partial \sigma _i$.

Equation of motion of $e$ is:
\be
-\partial L/\partial e= e^{-2}{\dot{x}}^2 +\sum_{i=1}^p {x,_i}^2=p^2+\sum_{i=1}^p {x,_i}^2=0 ,
\ee
\noindent and one can define $\lambda$ 's as below:
\be
\lambda _{m_1...m_p}=(1/\pi)^p\; \int_0^\pi d^p\sigma\;(p^2+\sum_{i=1}^p {x,_i}^2)
\; cos(m_1\sigma_1)...cos(m_p\sigma_p),
\ee

$$
m_i\in Z \kern -.5em Z^+\cup \{ 0\} ,
$$

Again the  Hamiltonian can be chosen as a linear combination of $\lambda$ 's and
we take:
\be
H=1/2\;\; \lambda_{0...0},
\ee
\noindent which leads to the equation of motion,
\be
\ddot{x}^\mu -\sum_{i=1}^p {{x^\mu},_i},_i=0
\ee

As in the case of strings, closure of the constraints algebra will give new 
constraints which must be added to the primary ones for the specification of 
the physical states\cite{15}.

\bigskip

\noindent{\large\bf Acknowledgement}
\medskip

\noindent It is a pleasure to thank H. Arfaei for helpful discussions.

\end{document}